\documentclass{PoS}

\usepackage{aas_macros}

\title{The Aid of Optical Studies in Understanding Millisecond Pulsar Binaries}

\ShortTitle{MSP Binaries}

\author{\speaker{Zorawar Wadiasingh}\\
        Centre for Space Research, North-West University, Potchefstroom, South Africa\\
        E-mail: \email{zwadiasingh [at] gmail.com}}

\author{Alice K. Harding \\
NASA Goddard Space Flight Center, 8800 Greenbelt Rd, MD 20771, United States \\
E-mail: \email{ahardingx [at] yahoo.com} }

\author{ Christo Venter \\
Centre for Space Research, North-West University, Potchefstroom, South Africa \\
E-mail: \email{christo.venter [at] nwu.ac.za} }

\author{ Markus Böttcher \thanks{NRF SARChI Chair} \\
Centre for Space Research, North-West University, Potchefstroom, South Africa\\
        E-mail: \email{Markus.Bottcher [at] nwu.ac.za }
}

\abstract{A large number of new ``black widow'' and ``redback'' energetic millisecond pulsars with irradiated stellar companions have been discovered through radio and optical searches of unidentified \textit{Fermi} sources. Synchrotron emission, from particles accelerated up to several TeV in the intrabinary shock, exhibits modulation at the binary orbital period. Our simulated double-peaked X-ray light curves modulated at the orbital period, produced by relativistic Doppler-boosting along the intrabinary shock, are found to qualitatively match those observed in many sources. In this model, redbacks and transitional pulsar systems where the double-peaked X-ray light curve is observed at inferior conjunction have intrinsically different shock geometry than other millisecond pulsar binaries where the light curve is centered at superior conjunction. We discuss, and advocate, how current and future optical observations may aid in constraining the emission geometry, intrabinary shock and the unknown physics of pulsar winds.}

\FullConference{SALT Science Conference 2015 -SSC2015-\\
		1-5 June, 2015\\
		Stellenbosch Institute of Advanced Study, South Africa}

\begin{document}

\vspace{-10mm}
\section{Background}

The \textit{Fermi} Large Area Telescope (LAT) has been prolific at discovering $\gamma$-ray pulsars, with millisecond pulsars (MSPs) representing a disproportionately high share of these detections.   Follow-up studies in the optical and radio of unassociated \textit{Fermi}-LAT sources and $\gamma$-ray MSPs have increased the number of known Galactic black widows (BWs) and redbacks (RBs) from four to almost 30, principally through precision radio timing of the MSPs revealing the existence of a companion by Doppler wobbling. These MSP binaries, first established with the discovery of B1957+20 \cite{1988Natur.333..237F}, are old $\gtrsim$ Gyr recycled systems with orbital periods $<1$ day in tight circular orbits. The pulsars are spun-up from their past accretion history and are now slowly ablating their companions by an energetic pulsar wind. 

The systems can be loosely grouped into two classes \cite{2011AIPC.1357..127R} based on the stellar companion's estimated mass $M_c$: BWs ($M_c \lesssim 0.05 M_\odot$) and RBs $M_c \gtrsim 0.1 M_\odot$. The RBs unusual companions are typically closer spectroscopically to the main sequence than those of BWs. In most of these systems, the energetics of emission in various wavebands are expected to be underpinned by the properties of the energetic pulsar wind, powered by electromagnetic rotational spin-down of the MSP. Crucially, relevant in the IR/optical context here, the secondary's luminosity may exceed, by several orders of magnitude, what one would naively expect from mass-luminosity relations for red dwarfs because of the irradiation and anisotropic heating of the photosphere by the $10^{34-35}$ erg s$^{-1}$ pulsar wind. Moreover, some BW companions may be white dwarfs. Recently, there has been some excitement over the discovery of transitional RB systems, e.g., J1023+0038 \cite{2009Sci...324.1411A} or J1824-2452I \cite{2013Natur.501..517P}, that switch between low-mass X-ray binary accretion-powered and radio pulsar rotation-powered states confirming the evolutionary link between these classes, with the accretion flow in the former state shrouding or switching off the pulsed radio emission from the pulsar. 

Our focus is on the rotation-powered state of BWs and RBs, where a relativistic intrabinary shock is expected to develop between the pulsar and companion \cite{1988Natur.333..832P}. Relativistic particle acceleration is expected to occur at this shock, where a large fraction of the pulsar spin-down power in a magnetically-dominated wind is converted particle energy and radiation \cite{1990ApJ...358..561H,1993ApJ...403..249A}. BWs and RBs are excellent astrophysical laboratories to study the unknown physics of pulsar winds and relativistic shock acceleration, probing the wind with a fixed target and photon field only $\sim 10^{11}$ cm away from the MSP position, unlike young pulsar wind nebulae where the termination shock may be $\sim 10^{15}$-$10^{17}$ cm away. The circular orbits, short orbital periods and multiple multiwavelength characterization of these systems offer considerable theoretical and observational advantages over other astrophysical systems for studying the physics of pulsar winds and relativistic shocks.

 Prolific X-ray emission modulated at the orbital period, likely synchrotron in origin, has been observed in several BWs and RBs e.g. \cite{2012ApJ...760...92H,2014ApJ...783...69G,2015arXiv150207208R}. Moreover, radio eclipses of the MSP, likely due to absorption, are observed centered around superior conjunction when the companion is between the pulsar and the observer.  The intrabinary shock is supported by either a companion wind or magnetosphere, and its geometry and location of the stagnation point where the respective wind pressures balance can significantly impact models of particle acceleration, cooling, and concomitant orbitally-modulated high-energy emission. Additionally, the orbital geometry and inclination can significantly impact the shrouding underpinning the radio eclipses, observed X-ray light curves and observability of $\gamma$-ray emission. We summarize how optical observations can service our understanding of these enigmatic systems.

\section{Our Model}

\begin{figure}[t]
\centering
\includegraphics[scale=0.265]{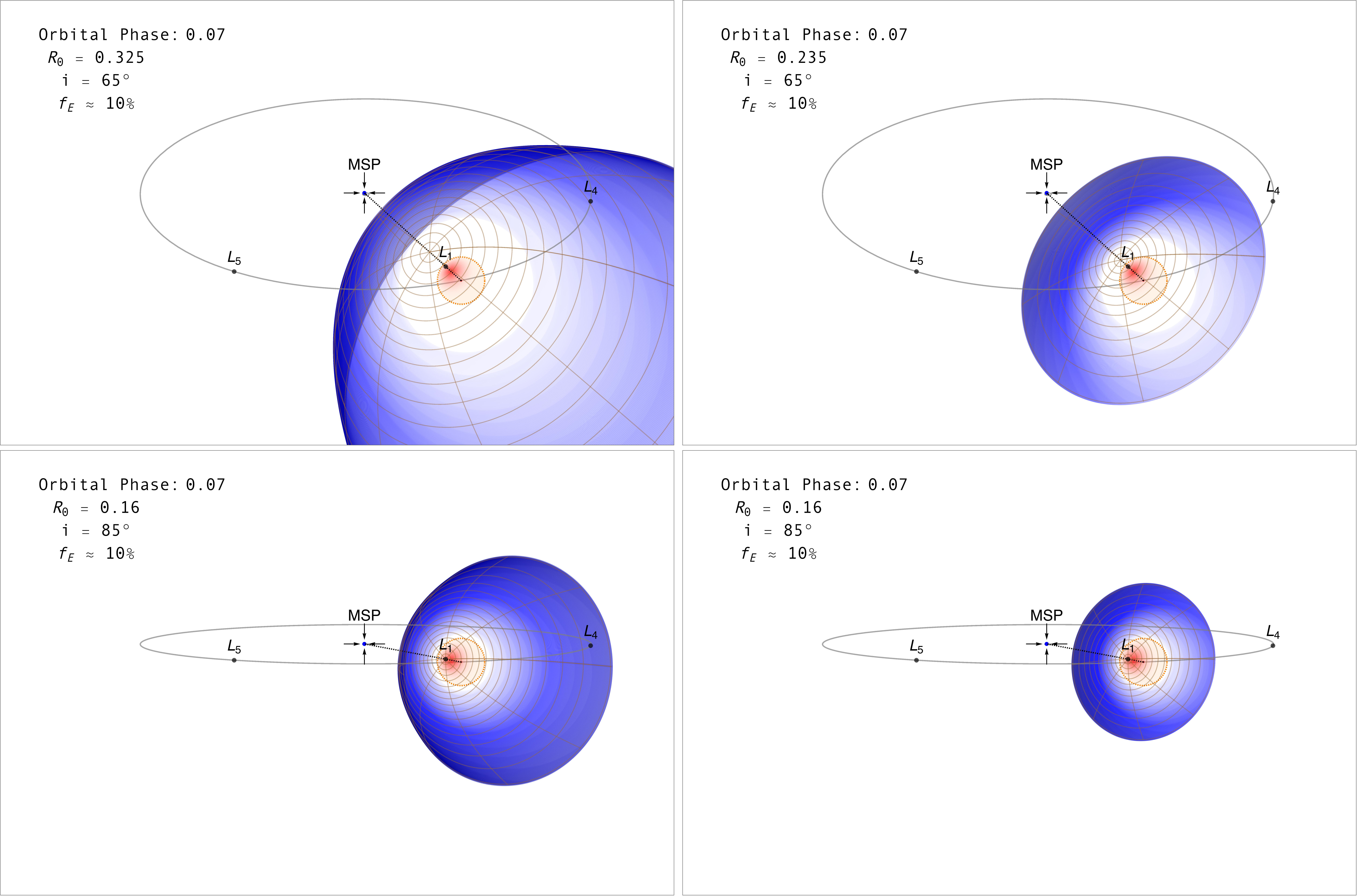}
 \vspace{-0.1truein}
\caption{Schematic representations of the PSR B1957+20 system, to scale, projected on the plane of the sky for inclinations $i = 65^\circ$ (upper panels) and $85^\circ$ (lower panels), mass ratio $69.2$ and normalized orbital phase $\approx 0.07$ from superior conjunction. The spherical companion is illustrated in red shading with an orange outline (the shading schematically representing the anisotropic heating by irradiation), for $90\%$ of the volumetric Roche radius \cite{1983ApJ...268..368E} with the black and blue dots representing the system barycenter and pulsar, respectively. Lagrange points are also depicted for clarity for the chosen mass ratio.  Left panels: a Type I shock scenario (see text for definition) is depicted near the companion for $R_0 = 0.325$ (in units of the orbit semimajor length, upper panel) and $0.16$ (lower panel) with a tail of length $3 R_0$ past the companion position outward from the pulsar position. Right panels: A Type II scheme with $R_0 = 0.235$ and $0.16$ extending to one-half the hemisphere of the heated companion. For both panels, the color coding emphasizes, schematically, the locales of first-order Doppler boosting. A velocity profile of $v_{\rm s} \propto \theta$ tangent to the shock is imposed; the blue indicates those regions of the shock where the $v_{\rm s}$ along the shock is toward the observer line-of-sight with intensity of coloring scaled with projected velocity component magnitude. }
\end{figure}  

We are in the process of developing a geometric model of the binary system and intrabinary shock, using the observed radio eclipses as a constraint on the spatial extent of the shock in an optically-thick formalism for radio absorption or scattering \cite{1989ApJ...342..934R}. We approximate the shocked pulsar wind and companion winds as being approximately spatially coincident, in a highly radiative limit, and azimuthally symmetric around the line joining the two masses. Analytic forms for shocks exist within this framework, for parallel-isotropic wind interaction \cite[Type I]{1996ApJ...459L..31W} and two colliding isotropic winds \cite[Type II]{1996ApJ...469..729C}.  There is then a one-to-one correspondence, for a fixed radio eclipse fraction, between the shock stand-off distance $R_0$ parameter, scaling the size of the shocked region, and system inclination for a prescribed shock geometry. Orbital sweepback of the shock is treated in our upcoming paper, resulting in asymmetry of the eclipses and light curves, although asymmetry in radio eclipses is only pronounced at the lowest observing frequencies. Figure~1 shows a diagram of the BW B1957+20 system for the symmetric shock case where the eclipse fraction is approximately $10\%$ for all four panels.

Many BWs and RBs show characteristic double-peaked light curves in the X-rays, either centered at superior or inferior conjunction. Although geometry shadowing by a bloated companion \cite{2011ApJ...742...97B} may explain some X-ray light curves, it cannot explain those double-peaked light curves centered around inferior conjunction in some RBs in the rotation powered-state, e.g. \cite{2010ApJ...722...88A, 2014ApJ...781L..21H,2015arXiv150207208R}. A natural explanation for such double-peaked emission is Doppler-boosted emission from mildly relativistic flow, likely a shocked pulsar wind rather than the nonrelativistic companion wind, along the intrabinary shock directed towards the observer. The shock may surround the companion or pulsar, the latter case not unexpected in transitional RB systems where an accretion-powered state may have occurred in the recent past. Our model is a work-in-progress, and we will soon incorporate more sophistication by modeling of particle transport along the shock, orbital sweepback, and assess the different components of inverse Compton emission from the various target photon fields present in these systems. We will also assess the influence of synchrotron self-absorption expected near the vicinity of the shock.
 

\section{How Future Optical Observations Can Help}

A cohesive understanding of many observational phenomena manifested in MSP binaries is still at an early stage of development. Our 3D modeling of MSP binaries' high-energy emission hinges upon from multiple observational inputs from other wavebands.  We advocate that IR/optical observations of BW and RB systems can constrain physics and geometry in several ways that has not yet been fully exploited. Only a small fraction of the currently known BWs and RBs have been studied in the IR/optical band, and even a smaller fraction with $8$-$10$ m class facilities such as SALT. Moreover, due to the dimness of some companions and other observational constraints, most have only been explored photometrically rather than spectroscopically typically with sparse cadence. We are also particularly interested in epochs of optical variability, discovered with either targeted monitoring campaigns or searches in wide-field surveys, that can be cross-correlated with >$8$ years of \textit{Fermi}-LAT data for $\gamma$-ray variability in these binaries.

\vspace{-2mm}
\paragraph{Discovery:} Optical follow-up of a \emph{Fermi} unidentified source 2FGL J1311.6-3429 revealed a periodic photometric light curve. The period determination constrained the generally intractable blind-search multidimensional parameter space, and enabled the discovery of the MSP from $\gamma$-ray data \cite{2012Sci...338.1314P} confirming the source's BW nature. A large number of other unassociated \emph{Fermi} sources have not yet been explored by optical facilities.

\vspace{-2mm}
\paragraph{Inclination and Masses:} 
Precision radio timing of the MSP yields the pulsar mass function, relating three variables: inclination, companion mass and neutron star mass. It is possible to construct physically-guided models of the anisotropic heating of the companion and fit the optical orbital flux modulation to constrain the inclination purely photometrically e.g. \cite{2007MNRAS.379.1117R, 2013ApJ...769..108B}. Deeper spectroscopic studies, assuming model atmospheres, can constrain radial velocities and the companion mass function. Combining the modeled inclination, pulsar mass function, and radial velocity measurement then fully determines both masses e.g. \cite{2011ApJ...728...95V}. It is found that BWs and RBs typically harbor heavy neutron stars $\gtrsim 2 M_\odot$ as expected from the recycling paradigm. The system inclination is especially critical for modeling observed orbitally-modulated X-ray and $\gamma$-ray emission from the intrabinary shock.

\vspace{-2mm}
\paragraph{Intrabinary Shock Constraints:}
If the intrabinary shock is supported by a companion magnetosphere rather than wind, then $\sim$ kG fields are required for typical MSP wind luminosities, to be consistent with the shock stand-off distances sufficient for the observed radio eclipses for a given inclination. Such fields have been measured in young magnetic stars by measuring Zeeman line broadening imprinted on the spectra, but the methods are rather involved \cite{2007ApJ...664..975J} and are limited to $\lesssim 12$-$14$ magnitudes for sufficient signal-to-noise with current-generation instruments. A future exploration of broadened spectral lines in the IR may constrain magnetic fields on some BWs or RBs and notably influence and advance our understanding of these systems, although a significant upgrade in sophistication may be necessary in model atmospheres allowing for the unusual ablated nature of the companions and anisotropic heating due to the pulsar wind. 

Depending on the plasma density near the companion controlling synchrotron self-absorption, an optical/UV nonthermal orbitally-modulated synchrotron component from the intrabinary shock may also be observable. Optical polarization may also be present if the mean large-scale magnetic field near the shock is well-ordered, although such an ordered field is not expected for high levels of mixing in the shock unless there is a significant companion magnetosphere influence. Such a detection, along with a model of particle transport will constrain the unknown relativistic electron population in the intrabinary shock and glean some information on the relativistic acceleration processes occurring in the shock.

\vspace{-2mm}
\paragraph{High-energy Emission:}
Some extreme BWs and RBs undergo profound optical flares \cite{2015ApJ...804..115R,2016arXiv160103681D}, dramatically increasing the target photon field for inverse Compton emission from either electrons in the intrabinary shock or the cold upstream pulsar wind. Such flares could be due to magnetic activity naturally generated by tidal interactions \cite{1992ApJ...385..621A}. Depending on the unknown particle distribution in the pulsar wind and downstream of the shock, such an orbitally-modulated inverse Compton component may be observable and give insight into the unshocked pulsar wind content.

%

\vspace{-1mm}
\acknowledgments
\vspace{-3mm}
The work of M.B. is supported by the South African Research Chairs Initiative of the Department of Science and Technology and the National Research Foundation of South Africa\footnote{Any opinion, finding and conclusion or recommendation expressed in this material is that of the authors and the NRF does not accept any liability in this regard.}. C.V. \& Z.W. are supported by the South African National Research Foundation. A.K.H. acknowledges support from the NASA Astrophysics Theory Program. A.K.H., Z.W., and C.V. also acknowledge support from the Fermi Guest Investigator Cycle 8 Grant.

\newcommand{\vol}[2]{$\,$\bf #1\rm , #2.}    
\def\mn{M.N.R.A.S.}
\def\aassupp{{Astron. Astrophys. Supp.}}
\def\apss{{Astr. Space Sci.}}
\def\apj{ApJ}
\def\nat{Nature}
\def\aaps{{Astron. \& Astr. Supp.}}
\def\aap{{A\&A}}
\def\apjs{{ApJS}}
\def\sp{{Solar Phys.}}
\def\jgr{{J. Geophys. Res.}}
\def\jphysb{{J. Phys. B}}
\def\ssr{{Space Science Rev.}}
\def\araa{{Ann. Rev. Astron. Astrophys.}}
\def\nature{{Nature}}
\def\asr{{Adv. Space. Res.}}
\def\rmp{{Rev. Mod. Phys.}}
\def\prc{{Phys. Rev. C}}
\def\prd{{Phys. Rev. D}}
\def\pr{{Phys. Rev.}}

\bibliographystyle{JHEP}
\bibliography{refs_BWRBs_SSC2015}

\end{document}